\documentclass[prd,superscriptaddress,psfig,nofootinbib]{revtex4}
\usepackage{amsmath}
\usepackage{amssymb}
\usepackage{amsfonts}
\usepackage{graphicx}
\usepackage{epsfig}
\usepackage{dcolumn}
\usepackage{hyperref}
\usepackage{epstopdf}
\usepackage{color}
\usepackage[utf8]{inputenc}
\usepackage{amsthm}
\usepackage{fontenc}
\usepackage{bm}
\RequirePackage{slashed}\usepackage{cancel}
\usepackage[normalem]{ulem}

\usepackage{array,booktabs}
\newcolumntype{C}{>{$\displaystyle}c<{$}}

\usepackage{soul}

\begin{document}


\title{Numerical evaluation of the nonlinear Gribov-Levin-Ryskin-Mueller-Qiu evolution equations for nuclear parton distribution functions}

\author{J. Rausch}
\affiliation{Humboldt University Berlin, Department of Physics, Newtonstra{\ss}e 15, 12489 Berlin, Germany}
\affiliation{Institut f\"ur Theoretische Physik, Westf\"alische Wilhelms-Universit\"at M\"unster, Wilhelm-Klemm-Stra{\ss}e 9, 48149 M\"unster, Germany}

\author{V. Guzey}
\affiliation{University of Jyvaskyla, Department of Physics, P.O. Box 35, FI-40014 University of Jyvaskyla, Finland}
\affiliation{Helsinki Institute of Physics, P.O. Box 64, FI-00014 University of Helsinki, Finland}

\author{M. Klasen}
\affiliation{Institut f\"ur Theoretische Physik, Westf\"alische Wilhelms-Universit\"at M\"unster, Wilhelm-Klemm-Stra{\ss}e 9, 48149 M\"unster, Germany}
\affiliation{School of Physics, The University of New South Wales, Sydney NSW 2052, Australia}

\date{\today}
\begin{abstract}

We numerically study for the first time the nonlinear GLR-MQ evolution equations for nuclear parton distribution function (nPDFs) to next-to-leading order accuracy and quantify the impact of gluon 
recombination at small $x$. Using the nCTEQ15 nPDFs as input, we confirm the importance of the nonlinear corrections for small 
$x \lesssim 10^{-3}$, whose magnitude increases with a decrease of $x$ and an increase of the atomic number $A$.
We find that at $x=10^{-5}$ and for heavy nuclei, after the upward evolution from $Q_0=2$ GeV to $Q=10$ GeV, the quark singlet 
$\Omega(x,Q^2)$ and the gluon $G(x,Q^2)$ distributions become reduced by $9-15$\%, respectively.
The relative effect is much stronger for the downward evolution from $Q_0=10$ GeV to $Q=2$ GeV, 
where we find that $\Omega(x,Q^2)$ is suppressed by 40\%, while $G(x, Q^2)$ is enhanced by 140\%. 
These trends propagate into the $F_2^A(x,Q^2)$ nuclear structure function and the $F_L^A(x,Q^2)$ longitudinal structure function,
which after the downward evolution become reduced by  45\% and enhanced by 80\%, respectively.
Our analysis indicates that the nonlinear effects are most pronounced in $F_L^A(x,Q^2)$ and are already quite sizable at
$x \sim 10^{-3}$ for heavy nuclei.
We have checked that our conclusions very weakly depend on the choice of input nPDFs. In particular, using the EPPS21 nPDFs as input, we
obtain quantitatively similar results.

\end{abstract}

\maketitle

\vspace*{-115mm} {\hfill MS-TP-22-48} \vspace*{100mm}

\section{Introduction}

In quantum chromodynamics (QCD), the microscopic structure of hadrons (pions, protons, nuclei) is described in terms of 
various quark and gluon (commonly called parton) distribution functions (PDFs). As follows from the QCD collinear factorization theorem~\cite{CTEQ:1993hwr}, the PDFs $f_i(x,Q^2)$ are universal, process-independent distributions, which depend on the 
parton flavor $i$, the parton light-cone momentum fraction of the parent hadron $x$, and the resolution scale $Q$. While the dependence
on $x$ cannot be calculated from first principles,   
the $Q^2$ dependence of $f_i(x,Q^2)$ is given by the Dokshitzer-Gribov-Lipatov-Altarelli-Parisi (DGLAP) evolution equations~\cite{Dokshitzer:1977sg,Gribov:1972ri,Gribov:1972rt,Altarelli:1977zs}.
In QCD and in any other quantum field theory with a dimensionless coupling constant, 
the $Q^2$ dependence of PDFs originates from renormalization of collinear divergences appearing in 
the ladder-type Feynman graphs (in the physical axial gauge) describing the emission of quarks and gluons with high transverse momenta (parton splitting)~\cite{Dokshitzer:1978hw}.
The resulting renormalization group equations are the DGLAP $Q^2$ evolution equations, which resum the leading 
$\alpha_s^k \alpha_s \ln Q^2$ contributions to these ladder graphs, where $k=0$ (leading-order of perturbation theory), $k=1$ (next-to-leading order, NLO), etc., 
and $\alpha_s(Q^2)$ is the QCD running coupling constant.

The standard DGLAP evolution equations have been derived in the limit of large $Q^2$ and $x \sim 1$ and are linear in the parton distributions.
The parton splitting encoded in these equations results in an increase of the quark and, especially, the gluon distributions at small $x$, when one increases the value of $Q^2$. When the gluon density becomes sufficiently large at small $x$, one needs to take into account the effects of gluon recombination (gluon-gluon fusion)
leading to nonlinear corrections to the DGLAP evolution equations~\cite{Gribov:1983ivg,Mueller:1985wy,Zhu:1999ht,Kovchegov:2012mbw}.
In the Gribov-Levin-Ryskin-Mueller-Qiu (GLR-MQ) approach~\cite{Gribov:1983ivg,Mueller:1985wy,Kovchegov:2012mbw}, the gluon recombination is addressed by analyzing so-called ``fan" diagrams, where two gluon ladders merge into a gluon or a quark-antiquark pair. 
Adding these contributions to the DGLAP equations yields the nonlinear GLR-MQ evolution equations~\cite{Mueller:1985wy,Qiu:1986wh},
where the nonlinear term tames the growth of the PDFs at small $x$ and leads to their suppression. This can be viewed as a precursor of
the gluon saturation at small $x$~\cite{Gelis:2010nm}.

Effects of small-$x$ nonlinear corrections to the DGLAP evolution equations due to gluon recombination have been extensively studied in the 
literature~\cite{Bartels:1990zk,Eskola:1993mb,Laenen:1995fh,Prytz:2001tb,Eskola:2002yc,Boroun:2009zzb,Boroun:2013mgv,Lalung:2018mpw,Devee:2014ida}.
It was found that these corrections affect the gluon distribution in the proton at small $x$,  $x \lesssim 10^{-3}$, and the interpretation 
and description of the Hadron-Electron Ring Accelerator (HERA) data on the total and diffractive electron-proton ($ep$) deep-inelastic scattering (DIS)
 cross sections at very small $x \sim 10^{-5}$. The effect of the nonlinear corrections is expected to be larger in heavy nuclei and also in models assuming the presence of gluonic ``hot spots" in the proton~\cite{Mantysaari:2016ykx}. This and many other topics of small-$x$ QCD constitute an essential part of the physics programs 
of future electron-ion colliders including the Electron-Ion Collider (EIC) in the U.S.~\cite{Accardi:2012qut}, 
the Large Hadron-Electron Collider (LHeC)~\cite{LHeCStudyGroup:2012zhm,Bruning:2019scy} and the Future Circular Collider (FCC)~\cite{FCC:2018byv} at CERN, which will allow one to access $ep$ DIS at as low as $x\sim 10^{-4}$ and $x\sim 10^{-6}$, respectively.

The aim of the present work is to study numerically for the first time the nonlinear corrections in the GLR-MQ evolution equations for nuclear parton distribution functions (nPDFs) to NLO accuracy. To this end, we extend the numerical algorithm
realized in the well-tested {\tt QCDNUM16} DGLAP evolution code~\cite{qcdnum} and write a stand-alone GLR-MQ evolution
program. 
As input, we use one of the state of the art nPDFs, namely the nCTEQ15 nPDFs~\cite{Kovarik:2015cma},  
which have been obtained by performing a global QCD fit of the data on lepton-nucleus DIS, Drell-Yan lepton pair production in proton-nucleus scattering at Fermilab, and inclusive pion production in deuteron-gold scattering at Relativistic Heavy Ion Collider (RHIC). 
We then solve the GLR-MQ equations numerically and 
quantify the effect of the nonlinear corrections in these equations on the evolved nPDFs and the nuclear structure function 
$F_2^A(x,Q^2)$ and the longitudinal structure function $F_L^A(x,Q^2)$. 
We find that, as expected, the nonlinear corrections are important for small $x \lesssim 10^{-3}$ and their magnitude increases with a decrease of $x$ and with an increase of the
 atomic number $A$. 
For the smallest studied value of $x=10^{-5}$, after the upward evolution from $Q_0=2$ GeV to $Q=10$ GeV, the quark singlet 
$\Omega(x,Q^2)$ and the gluon $G(x,Q^2)$ distributions in heavy nuclei are suppressed compared to their DGLAP-evolved counterparts by $9-15$\%, respectively. The relative effect is much stronger for the downward evolution from $Q_0=10$ GeV to $Q=2$ GeV, 
where we find that $\Omega(x,Q^2)$ is suppressed by 40\% compared to the nCTEQ15 PDFs, while $G(x, Q^2)$ is enhanced by 140\%. 
This trend can be explained by the observation that the gluon-gluon recombination plays a much bigger role than the gluon-quark
splitting. The behavior of nPDFs translates into the corresponding behavior of the $F_2^A(x,Q^2)$ and $F_L^A(x,Q^2)$
nuclear structure functions. In particular, after the downward evolution from high to low $Q$ and for heavy nuclei and very small $x$, we observe that $F_2^A(x,Q^2)$ dominated by $\Omega(x,Q^2)$ is reduced by 45\%, while $F_L^A(x,Q^2)$
dominated by $G(x,Q^2)$ is enhanced by 80\%.
We have also checked that these findings very weakly depend on the choice of input nPDFs and obtained quantitatively similar results
using the EPPS21 nPDFs~\cite{Eskola:2021nhw} as input.

The remainder of the paper is organized as follows.
In Sec.~\ref{sec:method}, we present our algorithm for the numerical solution of the DGLAP and GLR-MQ evolution equations.
The results of our numerical evaluation of the GLR-MQ equations for nPDFs and predictions for the $F_2^A(x,Q^2)$ and $F_L^A(x,Q^2)$ nuclear structure functions are given in Sec.~\ref{sec:results}. Finally, we summarize our findings 
in Sec.~\ref{sec:conclusions}.

\section{Numerical solution of GLR-MQ evolution equations}
\label{sec:method}

The standard DGLAP evolution equations have the following form for the singlet quark $\Omega(x,Q^2)=x\Sigma(x,Q^2)=x \sum_{i=u,d,s,c,\dots}(q_i(x,Q^2)+{\bar q}_i(x,Q^2))$ and the gluon  $G(x,Q^2)=xg(x,Q^2)$ momentum densities (distributions), 
\begin{eqnarray}
\frac{\partial \Omega(x,Q^2)}{\partial \ln Q^2} &=&\frac{\alpha_s(Q^2)}{2 \pi} \int^1_x \frac{dz}{z^2} x\left[P_{FF}\left(\frac{x}{z}\right) \Omega(z,Q^2)+
 P_{FG}\left(\frac{x}{z}\right) G(z,Q^2) \right]  \,, \nonumber\\
\frac{\partial G(x,Q^2)}{\partial \ln Q^2} &=&\frac{\alpha_s(Q^2)}{2 \pi} \int^1_x \frac{dz}{z^2} x\left[P_{GF}\left(\frac{x}{z}\right) \Omega(z,Q^2)+
 P_{GG}\left(\frac{x}{z}\right) G(z,Q^2) \right] \,,
\label{eq:DGLAP}	
\end{eqnarray}
where $P_{FF}$, $P_{FG}$, $P_{GF}$, and $P_{GG}$ are the quark-quark, gluon-quark, quark-gluon, and gluon-gluon splitting functions 
calculated to the desired order in $\alpha_s$~\cite{CTEQ:1993hwr,Furmanski:1980cm}. Our numerical analysis in this paper is carried out to NLO accuracy. 	

As we discussed in the Introduction, the gluon recombination modifies the standard DGLAP equations and leads to the following nonlinear GLR-MQ evolution equations~\cite{Mueller:1985wy,Qiu:1986wh}
\begin{eqnarray}
\frac{\partial \Omega(x,Q^2)}{\partial \ln Q^2} &=&\frac{\partial \Omega(x,Q^2)}{\partial \ln Q^2}\Big|_{{\rm DGLAP}}-\frac{27}{160}\frac{\alpha_s^2(Q^2)}{R^2Q^2}\big(G(x,Q^2)\big)^2 \,, \nonumber\\
\frac{\partial G(x,Q^2)}{\partial \ln Q^2} &=&\frac{\partial G(x,Q^2)}{\partial \ln Q^2}\Big|_{{\rm DGLAP}}
 -\frac{81}{16}\frac{\alpha_s^2(Q^2)}{R^2Q^2}\int_x^1\frac{dz}{z}\big(G(z,Q^2)\big)^2 \,,
\label{eq:GLRMQ_Qu}	
\end{eqnarray}
where $\partial \Omega(x,Q^2)/\partial \ln Q^2|_{\rm DGLAP}$ and $\partial G(x,Q^2)/\partial \ln Q^2|_{\rm DGLAP}$ refer
to the right-hand side of Eq.~(\ref{eq:DGLAP}). $R$ is the characteristic radius of the gluon distribution in
the hadronic target, which determines the strength of the nonlinear corrections.
%
Note that an additional term containing the higher-dimensional gluon distribution $G_\text{HT}$, which is suppressed by one power of $\ln 1/x$ 
and which does not correspond to the gluon distribution, has been neglected in Eq.~(\ref{eq:GLRMQ_Qu}). 
Since the non-singlet combinations of quark PDFs do not mix with with the gluon distribution and, hence, do not receive corrections due to gluon recombination, we do not consider them in our analysis.

We numerically solve the GLR-MQ evolution equations using the ``brute force" method in the momentum space.
To do it, we extend the numerical algorithm used in the {\tt QCDNUM16} DGLAP evolution code~\cite{qcdnum} to take into account the
nonlinear corrections in Eq.~(\ref{eq:GLRMQ_Qu}) and implement it in a stand-alone evolution code.
{\tt QCDNUM16} is a fast QCD evolution program, which numerically evolves PDFs using the DGLAP evolution equations to LO and NLO accuracy in the 
$\overline{\rm MS}$ factorization scheme. The program can handle flavor thresholds (light quark variable flavor number scheme or heavy quark fixed flavor number scheme) and allows one to independently vary the renormalization and factorization scales.
{\tt QCDNUM16} and its more recent variants constitute an important part of the open-source {\tt xFitter} project~\cite{Alekhin:2014irh,xFitter:2022zjb} 
providing a framework for the determination of the PDFs using QCD fits to the available data with lepton (HERA) and hadron (Tevatron, the Large Hadron Collider) beams.

Below we outline our approach.

Equations~(\ref{eq:DGLAP}) and (\ref{eq:GLRMQ_Qu}) are evaluated numerically on an $x-Q^2$ grid. 
Given the parton distributions at a starting value $Q_0^2$, the distributions at other values of 
$Q^2$ are determined by solving a set of four equations at each grid point, which are derived using spline interpolation between grid points.
The grid consists of $n+1$ values of $x$ bounded by $x_0$ and $1$, $x_0 < \dots < x_n =1$, and $m+1$ values of $Q^2$, which are all above or below 
$Q^2_0$. The values of $x$ and $Q^2$ are spaced logarithmically because the region of low $x$ and $Q^2$ is most relevant for our purpose. 
In the following, $D(x_c,Q^2_r) = D_{rc}$ refers to $\Omega(x,Q^2)$ or $G(x,Q^2)$ evaluated at the grid point $(x_c,Q^2_r)$, 
and the corresponding logarithmic derivative $\partial D/ \partial \ln Q^2$ is written as $D^{\prime}$. 
At $x=1$, $D_{rn} =0$ for all $r$.

To compute the convolution integrals in Eqs.~(\ref{eq:DGLAP}) and (\ref{eq:GLRMQ_Qu}), $D$ is interpolated linearly between $x$-values,
\begin{equation}
D_x(x,Q^2_r)=\bar{a}_1x+\bar{a}_0 \,,\ {\rm for}\ x \in [x_k,x_{k+1}].
\label{eq:m1}
\end{equation}
By imposing the continuity condition $D_x(x_i,Q^2_r)=D_{ri}$ for $i=k,k+1$, one obtains
\begin{equation}
D_x(x,Q^2_r)=(1-t_k)D_{rk} +t_k D_{r(k+1)}\,,\ {\rm where}\  t_k = \frac{x-x_k}{x_{k+1}-x_k} \in [0,1] \,.
\label{eq:m2}
\end{equation}
With this, the integrals can be written as weighted sums over $D_{rk}$, where $k$ runs from $c$ to $n$. Assuming $D_{rk}$ is known for $k > c$ (condition 1), the only unknowns in the DGLAP equations are $\Omega_{rc}$, $\Omega^{\prime}_{rc}$, $G_{rc}$, and $G^{\prime}_{rc}$.
Two more equations relating these unknowns can be obtained by using a quadratic interpolation between $Q^2$-values,
\begin{equation}
D_Q(x_c,Q^2)=\tilde{a}_2 (\ln Q^2)^2+\tilde{a}_1 \ln Q^2+\tilde{a}_0\,,\ {\rm for}\ Q^2 \in [Q^2_{r-1},Q^2_r] \,.
\label{eq:m3}
\end{equation}
The requirements that $D_Q(x_c,Q^2_i)=D_{ic}$ and $D_Q^{\prime}(x_c,Q^2_i)=D^{\prime}_{ic}$ for $i=r-1,r$ imply that
\begin{equation}
D_{rc} =D_{(r-1)c}+\frac{\Delta_r}{2}\left(D^{\prime}_{(r-1)c}+D^{\prime}_{rc}\right) \,, 
\label{eq:m4}
\end{equation}
where  $\Delta r = \ln Q^2_r-\ln Q^2_{r-1}$. If $D_{(r-1)c}$ and $D_{(r-1)c}^{\prime}$ have been calculated during the previous evolution steps (condition 2), Eq.~(\ref{eq:m4}) and the DGLAP equations can be solved for the four unknowns.

\begin{figure}[t]
\centering
 \epsfig{file=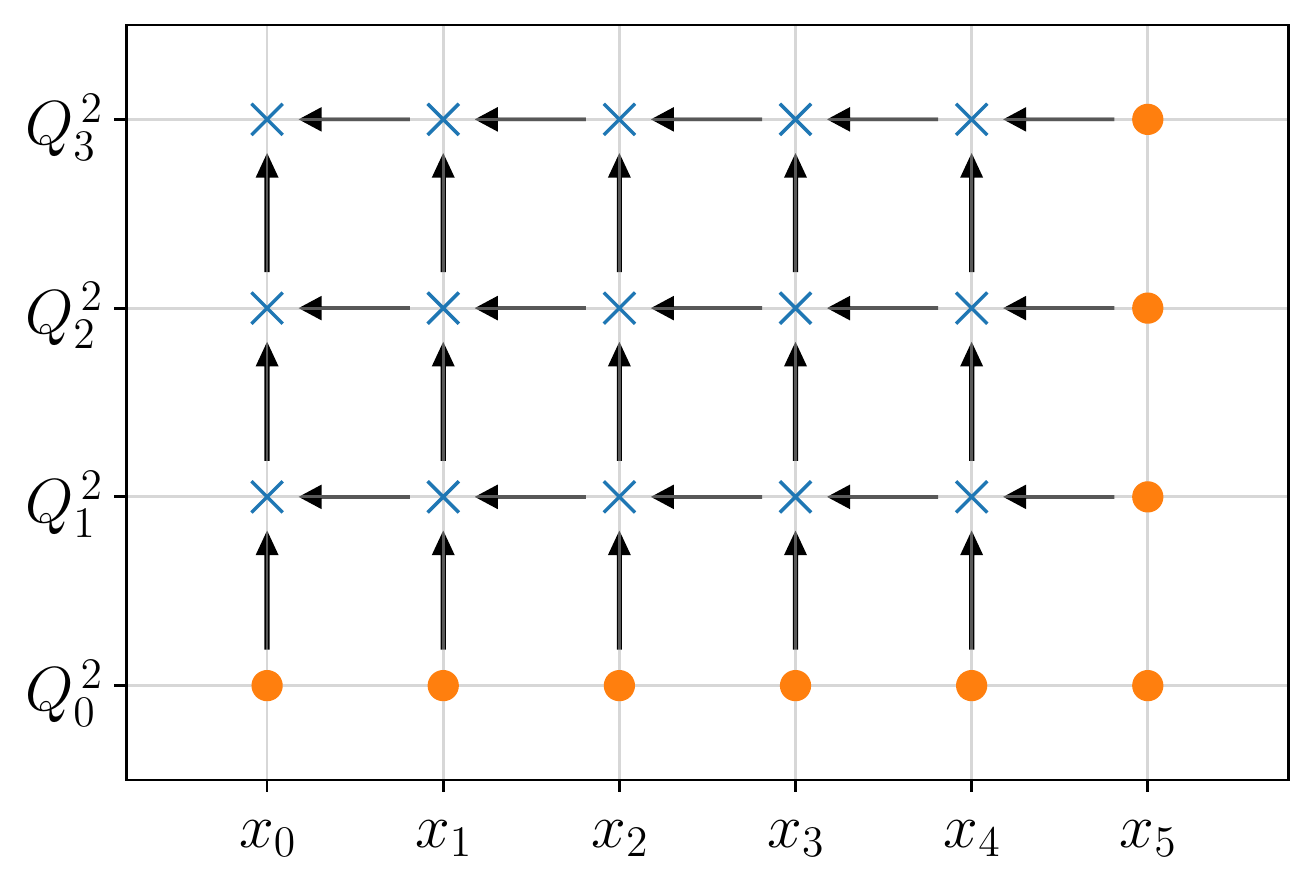,width=0.5\textwidth}
\caption{The evolution path through a grid with $n = 5$ and $m = 3$. Every row is evaluated from the right to the left, starting at the bottom and going to the top. The orange dots indicate the starting values of $D$.}
\label{fig:path}
\end{figure}

The path through the grid must now be chosen such that conditions 1 and 2 are satisfied for any grid point $(x_c,Q^2_r)$, where $D$ is being evaluated. This is achieved by starting at $x_n = 1$ for every value of $Q^2$ and proceeding towards smaller $x$, as illustrated in Fig.~\ref{fig:path}.

Using the linear interpolation of Eq.~(\ref{eq:m2}), the convolution integrals in the DGLAP equations can be written as
\begin{equation}
\int_{x_c}^1 \frac{dz}{z^2} x_c P_{AB}\left(\frac{x_c}{z}\right) D(z,Q^2_r)=\sum_{k=c}^{n-1} \omega_{AB}(x_k,x_c) D_{rk} \,,
\label{eq:m5} 
\end{equation}
where
\begin{equation}
\omega_{AB}(x_k,x_c)=\left\{\begin{array}{ll}
S_1(f_{c+1},f_c) & {\rm if}\ k=c \\
S_1(f_{k+1},f_k)-S_2(f_k,f_{k-1}) & {\rm else} 
\end{array} \right.
\label{eq:m6} 
\end{equation}
with $f_k = x_c/x_k$ and
\begin{equation}
S_i(u,v)=\frac{a_i}{v-u} \int^{v}_{u} \frac{dz}{z} (z-b_i) P_{AB}(z) \,,
\label{eq:m7}
\end{equation}
where $a_1=b_2=v$ and $a_2=b_1=u$. The weights $w_{AB}(x_k,x_c)$ are calculated numerically at program initialization.

The discretized DGLAP equations can then be expressed in the following form
\begin{eqnarray}
\Omega^{\prime}_{rc} &=& W_{FF} \Omega_{rc} + W_{FG} G_{rc} + M_F \,, \nonumber\\ 
G^{\prime}_{rc} &=& W_{GF} \Omega_{rc} + W_{GG} G_{rc} + M_G \,,
\label{eq:m8}
\end{eqnarray}
where $W_{AB} = \alpha_s/(2 \pi))w_{AB}(x_c,x_c)$ and $M_F$ and $M_G$ contain the summands with $k > c$ 
multiplied by $\alpha_s/(2 \pi)$. Together with Eq.~(\ref{eq:m4}), they form a system of four linear equations with four unknowns. This system is solved numerically at every step in the evolution.

The nonlinear correction in the GLR-MQ evolution equations involves the gluon distribution squared, see Eq.~(\ref{eq:GLRMQ_Qu}).
The difficulty presents only the second line in Eq.~(\ref{eq:GLRMQ_Qu}), where the nonlinear term is expressed as an integral  
of $(G(z,Q^2))^2$.
To numerically implement it, one uses the linear interpolation in $z$ for $G(z,Q^2)$, see Eqs.~(\ref{eq:m1}) and (\ref{eq:m2}), 
substitutes it in Eq.~(\ref{eq:GLRMQ_Qu}), and obtains the following discretized expression for the nonlinear correction,
\begin{equation}
\int^1_{x_c} \frac{dz}{z} G^2(z,Q^2_r)=\sum_{k=c}^{n-1} w_1(x_k)G^2_{rk}+\sum_{k=c}^{n-2} 2w_2(x_k)G_{r(k+1)}G_{rk} \,,
\label{eq:m9}
\end{equation}
where
\begin{eqnarray}
w_1(x_k) &=& \left\{\begin{array}{ll}
T_1(x_c, x_{c+1}) & {\rm if}\ k = c \\
T_1(x_k, x_{k+1}) + T_2(x_{k_1},x_k) & {\rm else}
\end{array} \right. \nonumber\\
w_2(x_k) &=& -T_3(x_k, x_{k+1}) 
\label{eq:m10}
\end{eqnarray}
and 
\begin{equation}
T_i(u,v)=\frac{1}{(v-u)^2}\int^{v}_{u} \frac{dz}{z} (c_i d_i -(c_i+d_i)z+z^2) \,,
\label{eq:m11}
\end{equation}
where $c_1=d_1=d_3 =v$ and $c_2=d_2=c_3 =u$.
Note that since the nonlinear correction is expressed in terms of the gluon distribution squared, Eq.~(\ref{eq:m9}) contains
both the $G^2_{rk}$ terms and the $G_{r(k+1)}G_{rk}$ cross terms [compare to Eq.~(\ref{eq:m5})].

The computation of $w_1(x_k)$ and $w_2(x_k)$ can be done much faster than that of the DGLAP weights $w_{AB}(x_k,x_c)$: 
Only ${\cal O}(n)$ integrals must be calculated instead of ${\cal O}(n^2)$, and the integrand in Eq.~(\ref{eq:m11}) is much simpler 
than the splitting functions in Eq.~(\ref{eq:m5}).

With this, the discretized form of the GLR-MQ evolution equations read
\begin{eqnarray}
\Omega^{\prime}_{rc} &=& W_{FF} \Omega_{rc} -V_1 G^2_{rc}+ W_{FG} G_{rc} + M_F \,, \nonumber\\ 
G^{\prime}_{rc} &=& W_{GF}\Omega_{rc} -V_2 G^2_{rc}+(W_{GG}-V_3) G_{rc} + M_G-N_G \,,
\label{eq:m12}
\end{eqnarray}
where $V_1 = (27/160)f(Q^2_r)$, $V_2 = (81/16)f(Q^2_r)w_1(x_c)$, $V_3 = (81/16)f(Q^2_r)2w_2(x_c)G_{r(c+1)}$, and 
$f(Q_r^2)=\alpha_s^2(Q_r^2)/(R^2 Q_r^2)$. 
The $N_G$ term contains the remainder of the sums in Eq.~(\ref{eq:m9}) multiplied by the factor of $(81/16)f(Q^2_r)$. 
Since Eq.~(\ref{eq:m4}) still applies, there are again four equations relating $D_{rc}$ and $D_{rc}^{\prime}$, 
which can be solved at each grid point, when using the evolution path shown in Fig.~\ref{fig:path}.

Using the numerical approach outlined above, we solved the GLR-MQ evolution equations on a $50 \times 40$ grid
($n = 50$, $m = 40$) in the $x-Q^2$ plane
using the nCTEQ15~\cite{Kovarik:2015cma}
and EPPS21~\cite{Eskola:2021nhw} nPDFs
 for the initial condition,
 which
 have been accessed via the 
LHAPDF6 framework~\cite{Buckley:2014ana}. 
This grid size is sufficient to provide a better than 1.2\% numerical accuracy for the interpolation in $x$ and $Q^2$.

For the running strong coupling constant $\alpha_s(Q^2)$, 
we used the standard NLO expression~\cite{CTEQ:1993hwr} along with the requirements that $\alpha_s(M_Z^2)=0.118$, where $M_Z=91.2$ GeV is the $Z$ boson mass, and that $\alpha_s(Q^2)$ is continuous across the charm quark mass $m_c=1.3$ GeV and the bottom quark mass $m_b=4.5$ GeV 
flavor thresholds.

For a nuclear target with the mass number $A$, we take $R=1.25\ {\rm fm}\times A^{1/3}$.
Note that since nuclear PDFs scale approximately as $A$, the nonlinear term in Eq.~(\ref{eq:GLRMQ_Qu}) scales as $A^{4/3}$, which 
significantly enhances the importance of the nonlinear corrections for heavy nuclei compared to the proton case. However, in practice,
the significant nuclear shadowing of the gluon distribution at small $x$ and the rather dilute distribution of nucleons in nuclei
reduce the net effect~\cite{Frankfurt:2022jns}.

To test the accuracy of our evolution code, as an example, we used the nCTEQ15 nPDFs for Au-197 as the initial conditions at $Q_0=2$ GeV, evolved them
up to $Q=10$ GeV neglecting the nonlinear GLR-MQ correction, and found that the resulting quark singlet and gluon distributions in the
$10^{-5} \leq x \leq 10^{-3}$ interval  agree with the nCTEQ15 parametrization with an accuracy of around 1.2\%.

\section{Results for nuclear PDFs and structure functions}
\label{sec:results}

\begin{figure}[t]
\centering
\epsfig{file=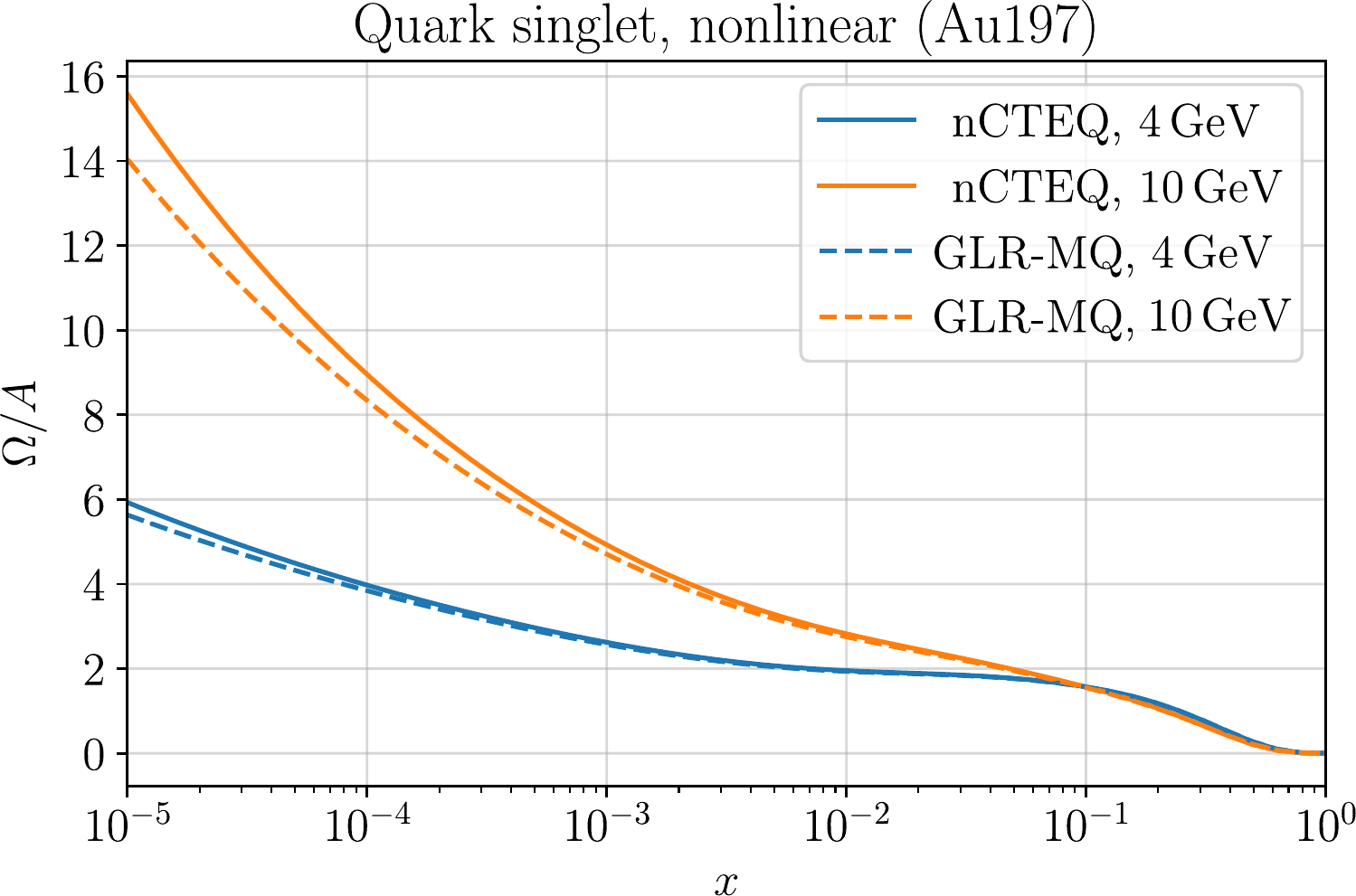,width=0.475\textwidth}
\epsfig{file=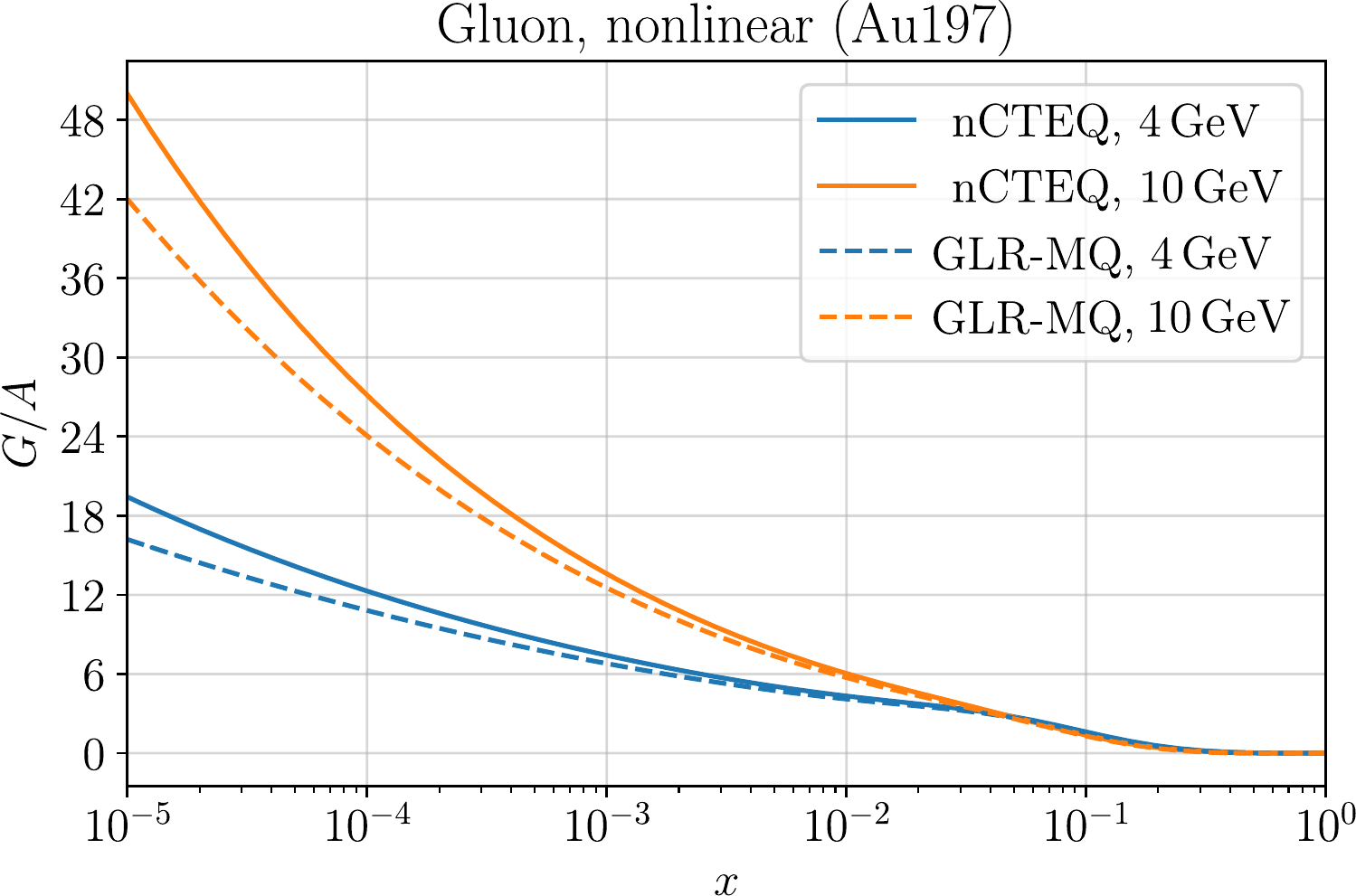,width=0.475\textwidth}
\epsfig{file=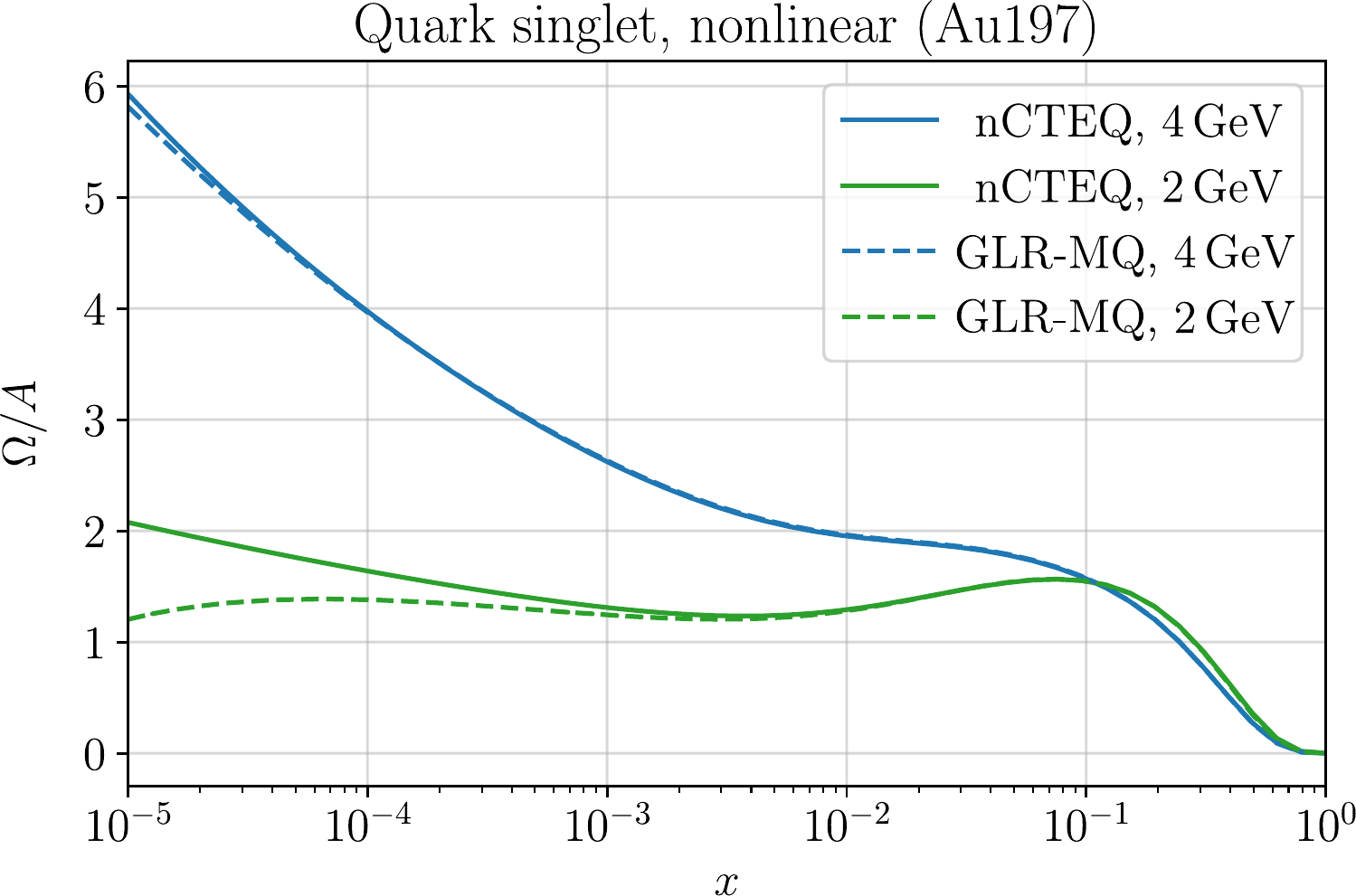,width=0.475\textwidth}
\epsfig{file=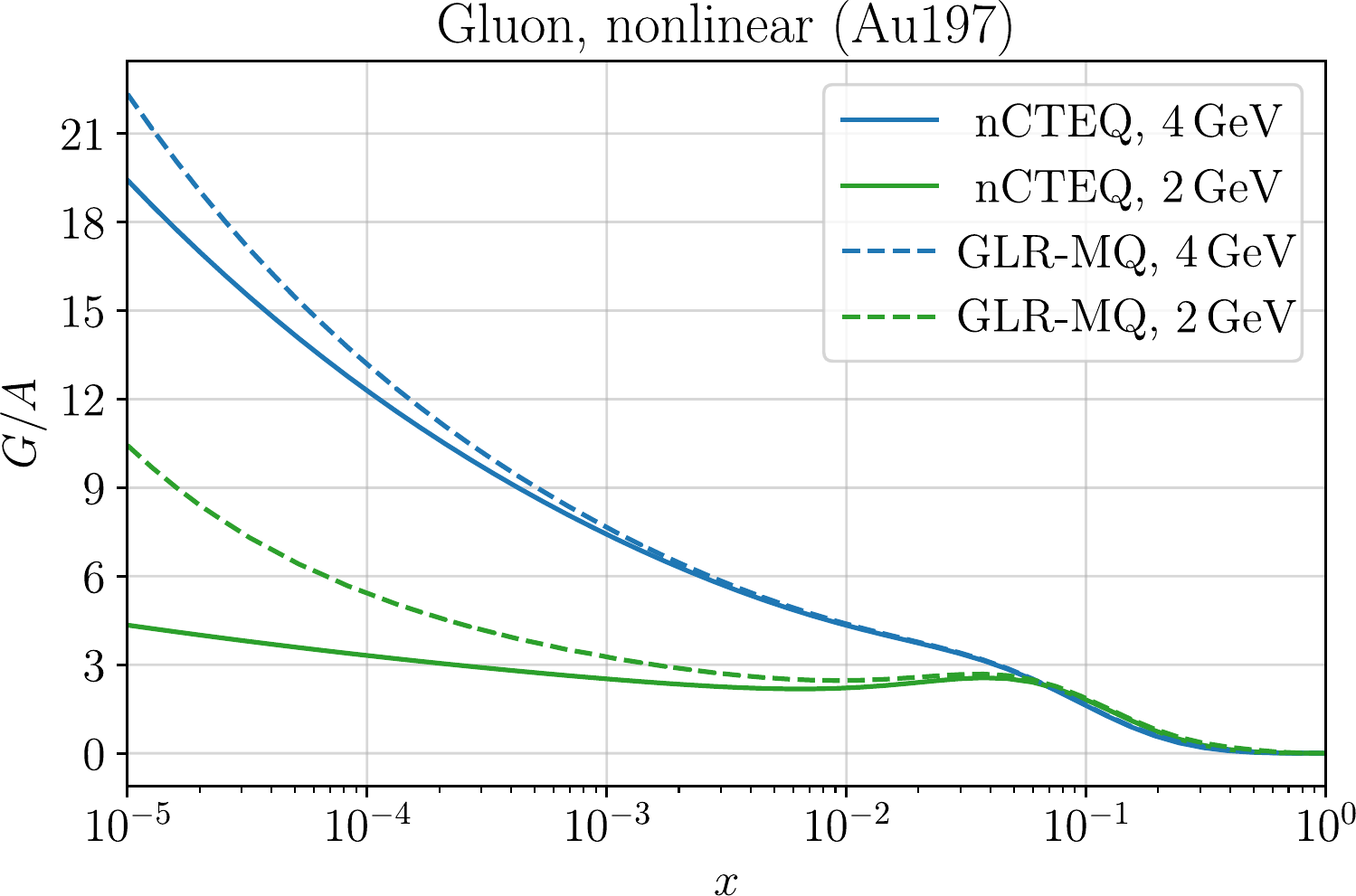,width=0.475\textwidth}
\caption{Results of the nonlinear GLR-MQ evolution equations for nPDFs of the nucleus of Au-197.
The quark singlet $\Omega(x,Q^2)$ and the gluon $G(x,Q^2)$ distributions per nucleon (dashed lines) are shown as a function of $x$ 
after the upward evolution from $Q_0=2$ GeV to $Q=4$ GeV and $Q=10$ GeV (two upper panels) and
after the downward evolution from $Q_0=10$ GeV to $Q=4$ GeV and $Q=2$ GeV (two lower panels) using the nCTEQ15 input.
For comparison, the solid curves show the results of the nCTEQ parametrization at the corresponding values of $Q$.
}
\label{fig:nonlin_up_down}
\end{figure}

In this section, we present results of our numerical studies of the nonlinear GLR-MQ evolution equations for the nCTEQ15 
and EPPS21
 nPDFs,
quantify the effect of the nonlinear corrections in these equations on the evolved nPDFs and the nuclear structure function $F_2^A(x,Q^2)$ and
the longitudinal structure function $F_L^A(x,Q^2)$, and thus determine the kinematic regions in the $x-Q^2$ plane, where the nonlinear 
corrections are potentially important.

Figure~\ref{fig:nonlin_up_down} presents the results of GLR-MQ evolution for the quark singlet (left panels) and the gluon (right
panels) nPDFs divided by $A$ for the heavy nucleus of Au-197 as a function of the momentum fraction $x$. 
In the two upper panels, the dashed curves labeled ``GLR-MQ" show the results the upward evolution from $Q_0=2$ GeV to $Q=4$ GeV and $Q=10$ GeV.
They are compared to the nCTEQ15 parametrization at the corresponding values of $Q$ given by the solid curves. As expected, 
the recombination of low-$x$ gluons into high-$x$ gluons slows down the growth of both gluon and quark singlet distributions 
for $x < 10^{-3}$.  
Since the gluon-quark splitting function $P_{FG}(x/z)$ is positive for $x \leq z \leq 1$, the lower gluon PDFs lead to a smaller rate of 
change $\Omega^{\prime}(x,Q^2)$ and, consequently, to the observed decrease in $\Omega(x,Q^2)$. 
The absolute difference between the GLR-MQ and DGLAP evolved PDFs is generally smaller for the quark distribution.
For instance, the values of $\Omega(x,Q^2)$ and $G(x,Q^2)$
at $Q=10$ GeV and $x=10^{-5}$ are reduced by 10\% and 14\%, respectively, compared to the corresponding nCTEQ15 PDFs.

The nonlinear terms in the GLR-MQ equations are suppressed as $1/Q^2$ and, hence, evolving downwards from a high value of $Q_0$ should 
in principle give a more accurate picture of the importance of the gluon recombination effect due to the evolution because the input is now 
insignificantly affected by gluon recombination. This is presented in the two lower panels of Fig.~\ref{fig:nonlin_up_down}
showing by the dashed curves the results of the downward evolution from $Q_0=10$ GeV down to $Q=4$ GeV and $Q=2$ GeV. 
The relative deviation from the nCTEQ15 parametrization given by the solid curves is notably larger than for the upward evolution
because the quark and gluon distributions are much smaller at low $Q$.
 For instance, at $Q=2$ GeV and $x=10^{-5}$, $\Omega(x,Q^2)$ is decreased by 43\% compared to the nCTEQ15 PDFs, while $G(x, Q^2)$ is increased by 133\%. 
 At the same time, the absolute difference between the GLR-MQ and DGLAP evolved PDFs is similar for both evolution directions.
 
\begin{figure}[t]
\centering
\epsfig{file=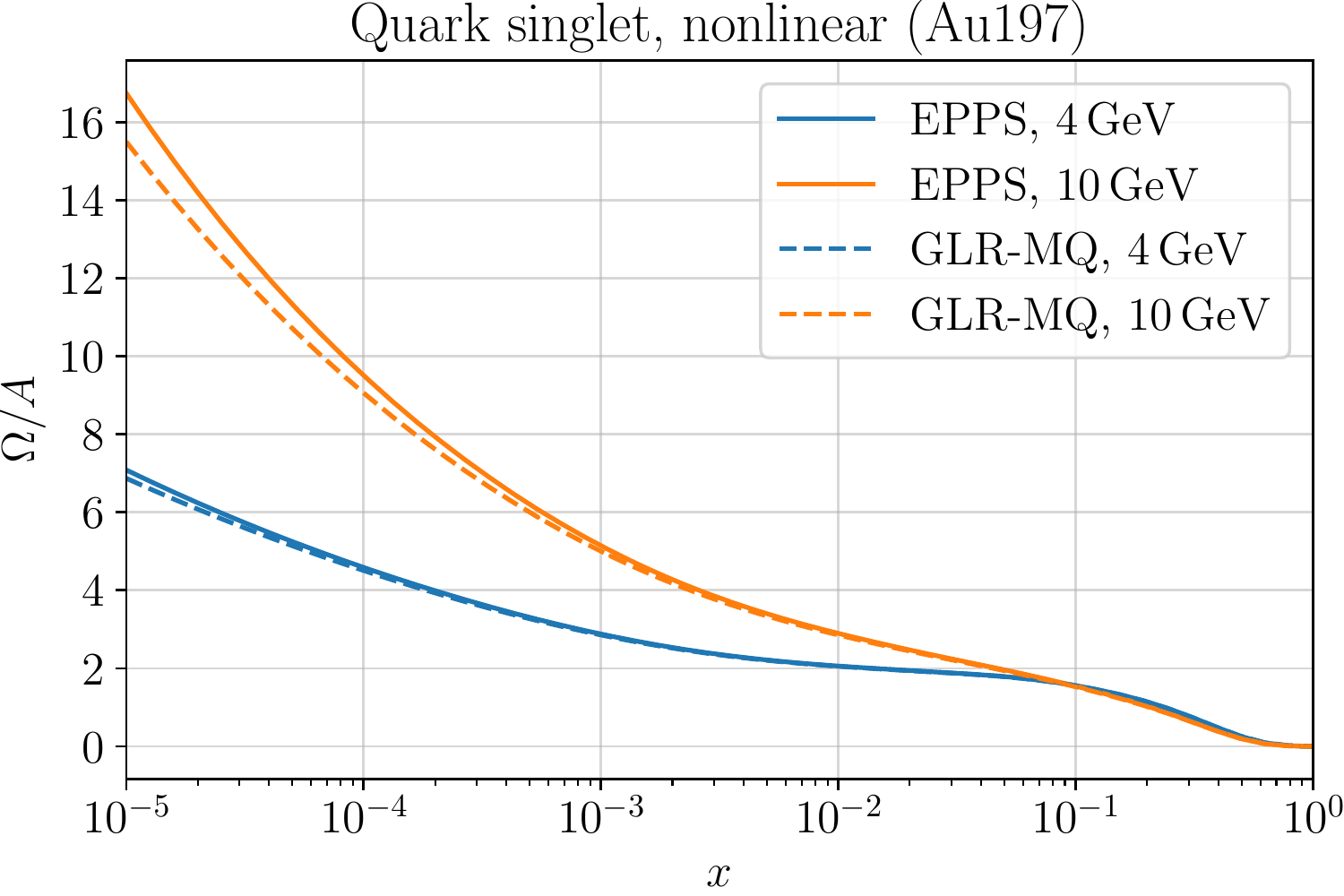,width=0.475\textwidth}
\epsfig{file=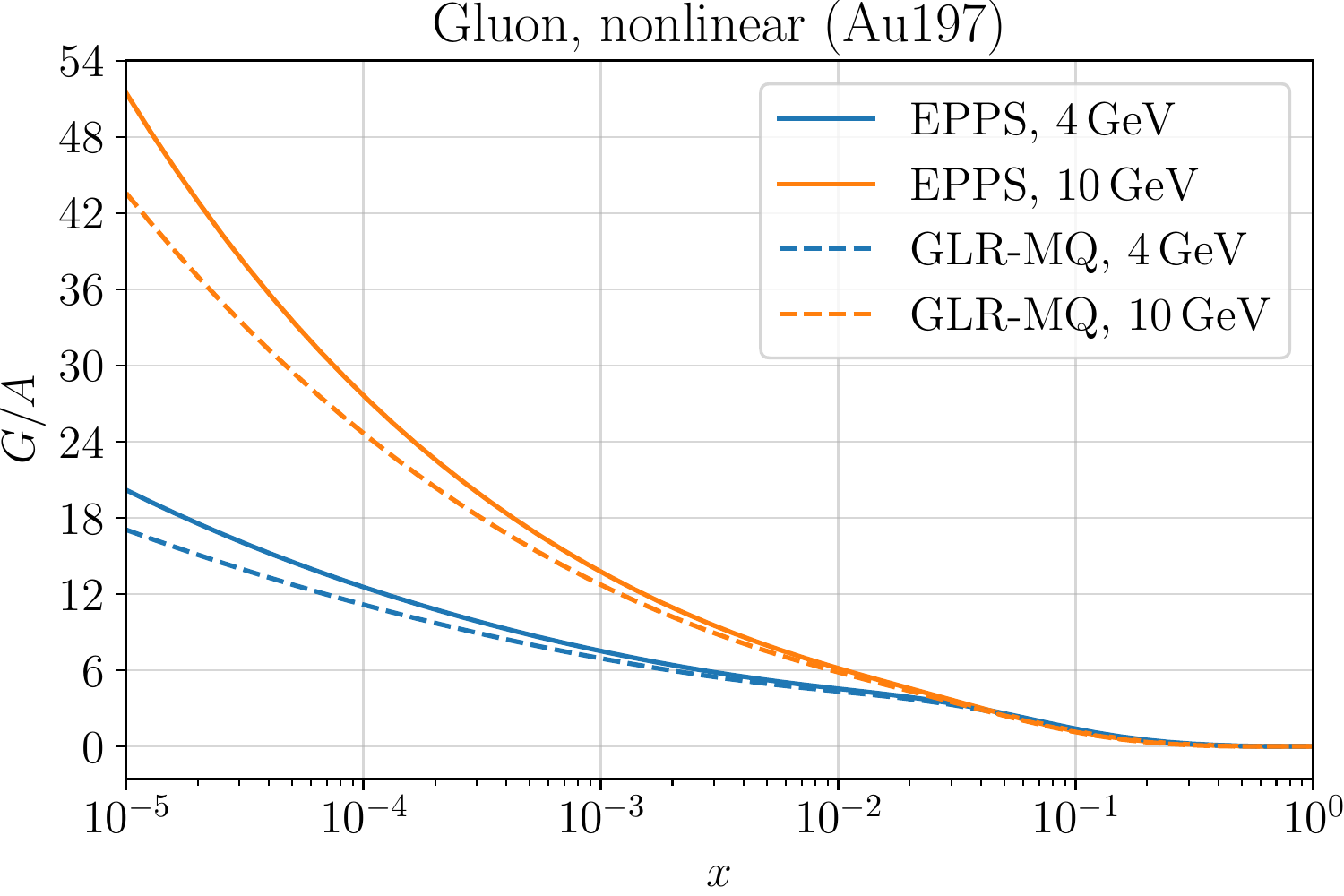,width=0.475\textwidth}
\epsfig{file=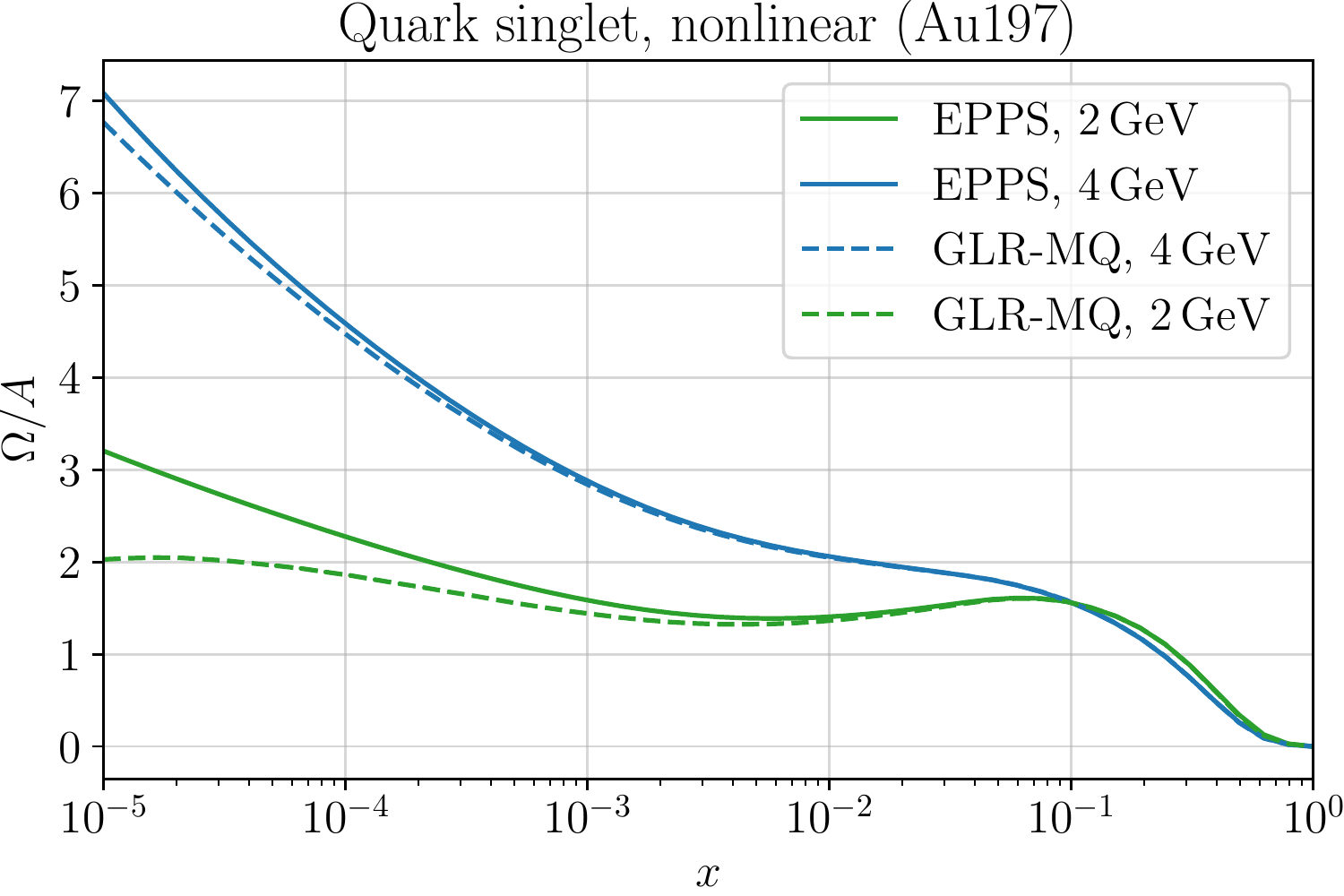,width=0.475\textwidth}
\epsfig{file=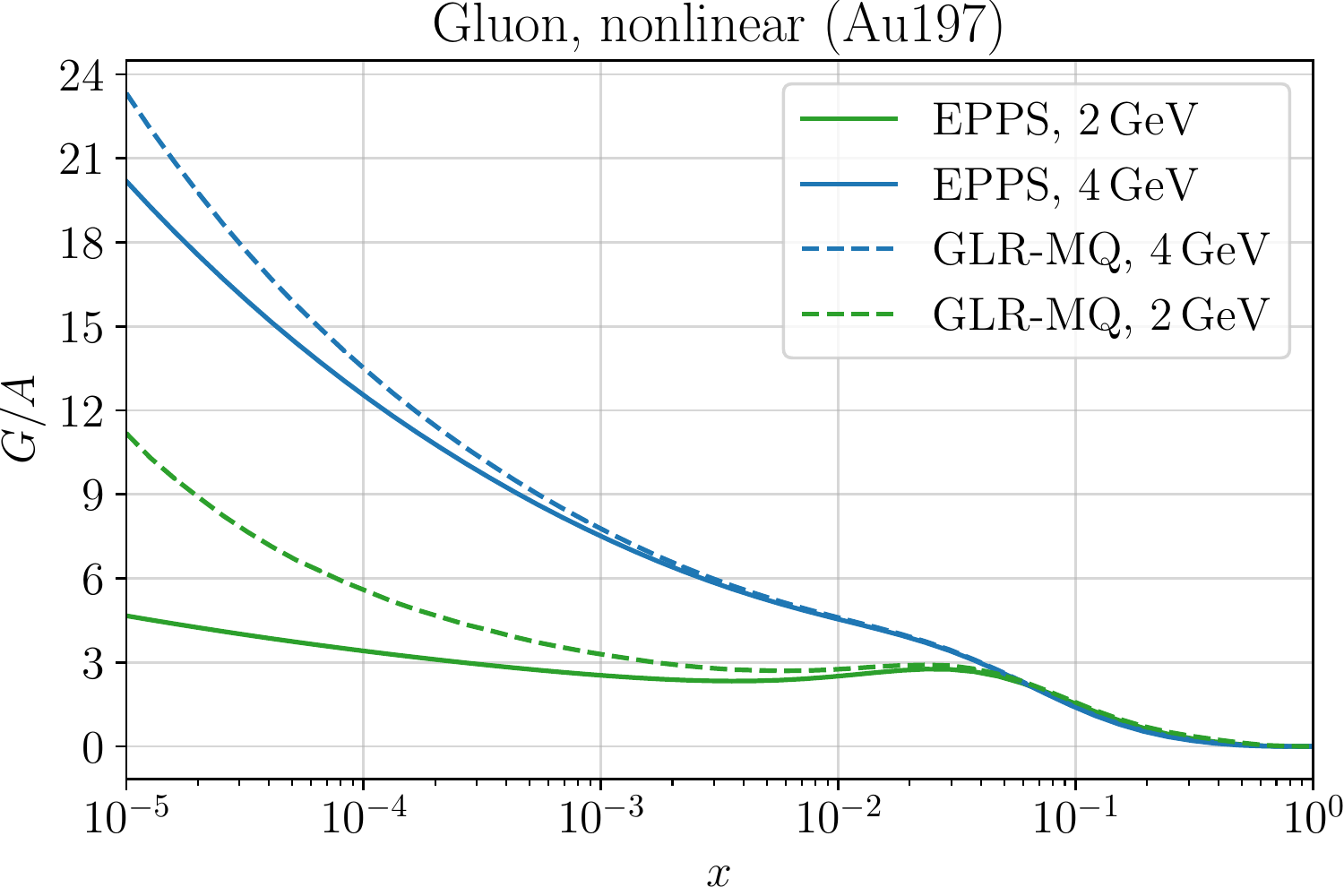,width=0.475\textwidth}
\caption{The same as Fig.~\ref{fig:nonlin_up_down}, but with the EPPS21 nPDFs.}
\label{fig:EPPS_nonlin_up_down}
\end{figure}

The gluon distribution after the downward evolution can be seen as a consequence of reversed gluon-gluon recombination, i.e., 
the migration of high-$x$ gluons towards low $x$ leading to the observed increase. As in the case of the upward evolution, 
the change in $G(x,Q^2)$ affects $\Omega(x,Q^2)$ mostly through the $P_{FG}(x/z)$ splitting function. 
The gluon-quark splitting in the case of the downward evolution corresponds to quark-antiquark pairs recombining into gluons, which explains 
the decrease in $\Omega(x, Q^2)$ observed in the lower left panel of Fig.~\ref{fig:nonlin_up_down}.

To better understand and isolate the role of the gluon recombination effects in the evolution of quarks, we also
solved the GLR-MQ equations without parton mixing by setting $P_{FG}=P_{GF}=0$. 
We observed that this essentially stops the $Q^2$ evolution of the quark singlet distribution, which indicates that 
the combined effect of the quark-quark splitting and the gluon-quark recombination is very small compared to that of the neglected gluon-quark 
splitting. Consequently, the differences between the GLR-MQ and the DGLAP evolved quark PDFs can mainly be attributed to the $g-g$ recombination. 
This is consistent with the predictions made by Mueller and Qiu~\cite{Mueller:1985wy}.

We also repeated our analysis using the EPPS21 nPDFs as input; the corresponding results are presented in Fig.~\ref{fig:EPPS_nonlin_up_down}.
A comparison of these results to those in Fig.~\ref{fig:nonlin_up_down} shows that they are quantitatively very similar.
Therefore, our conclusions on the trends and magnitudes of the nonlinear effects in the GLR-MQ equations very weakly depend on the choice of input nPDFs. 

\begin{figure}[t]
\centering
\epsfig{file=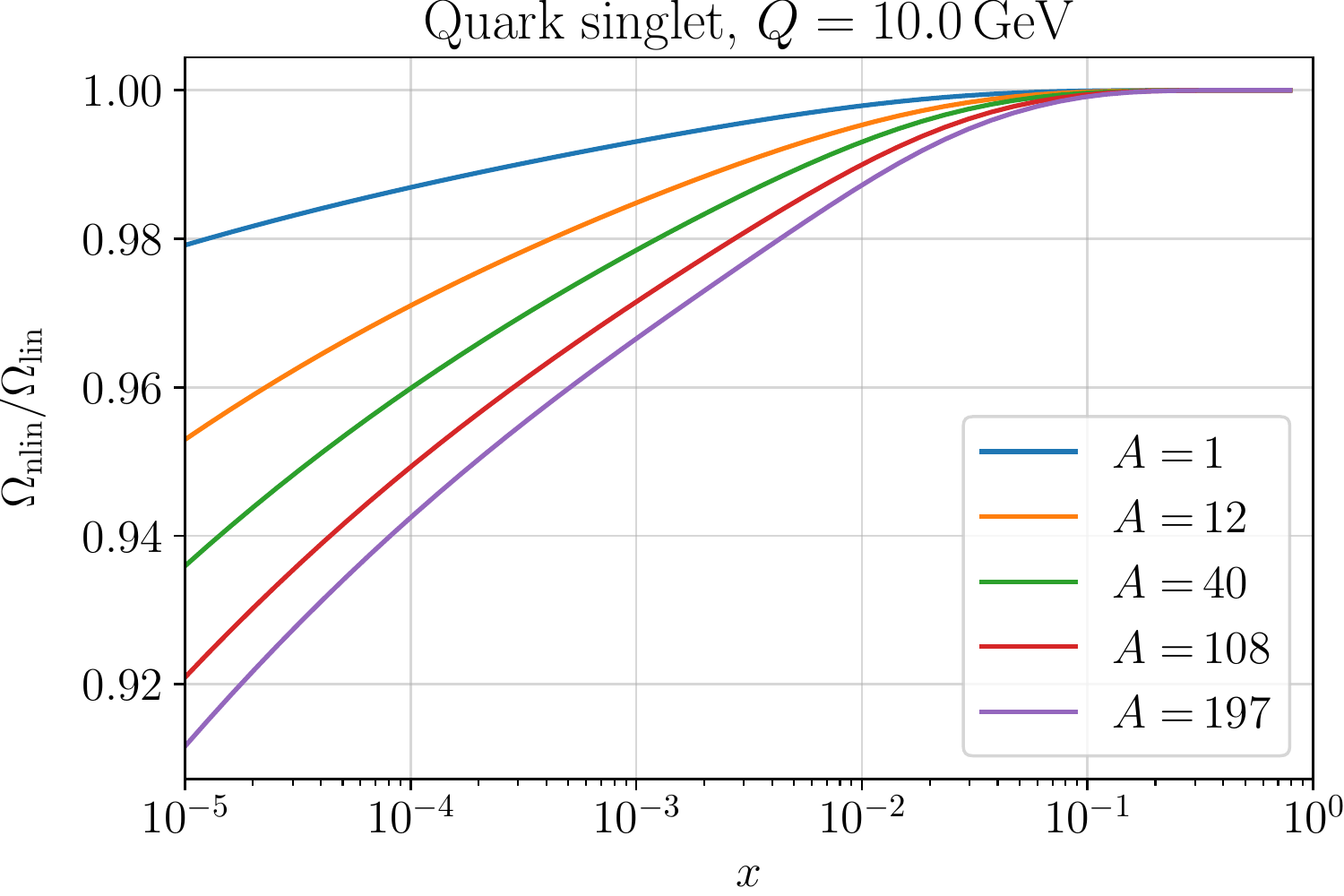,width=0.475\textwidth}
\epsfig{file=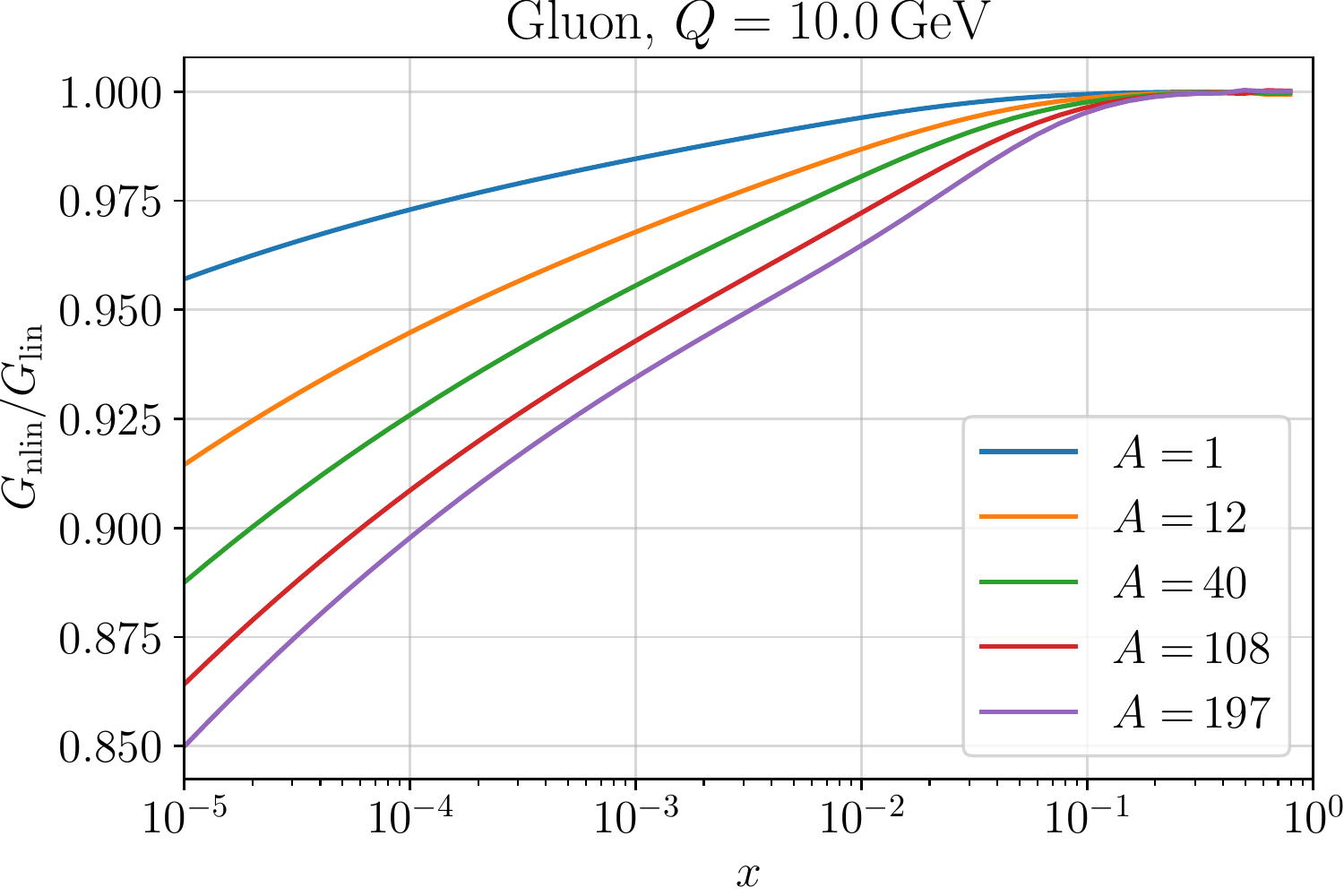,width=0.475\textwidth}
\epsfig{file=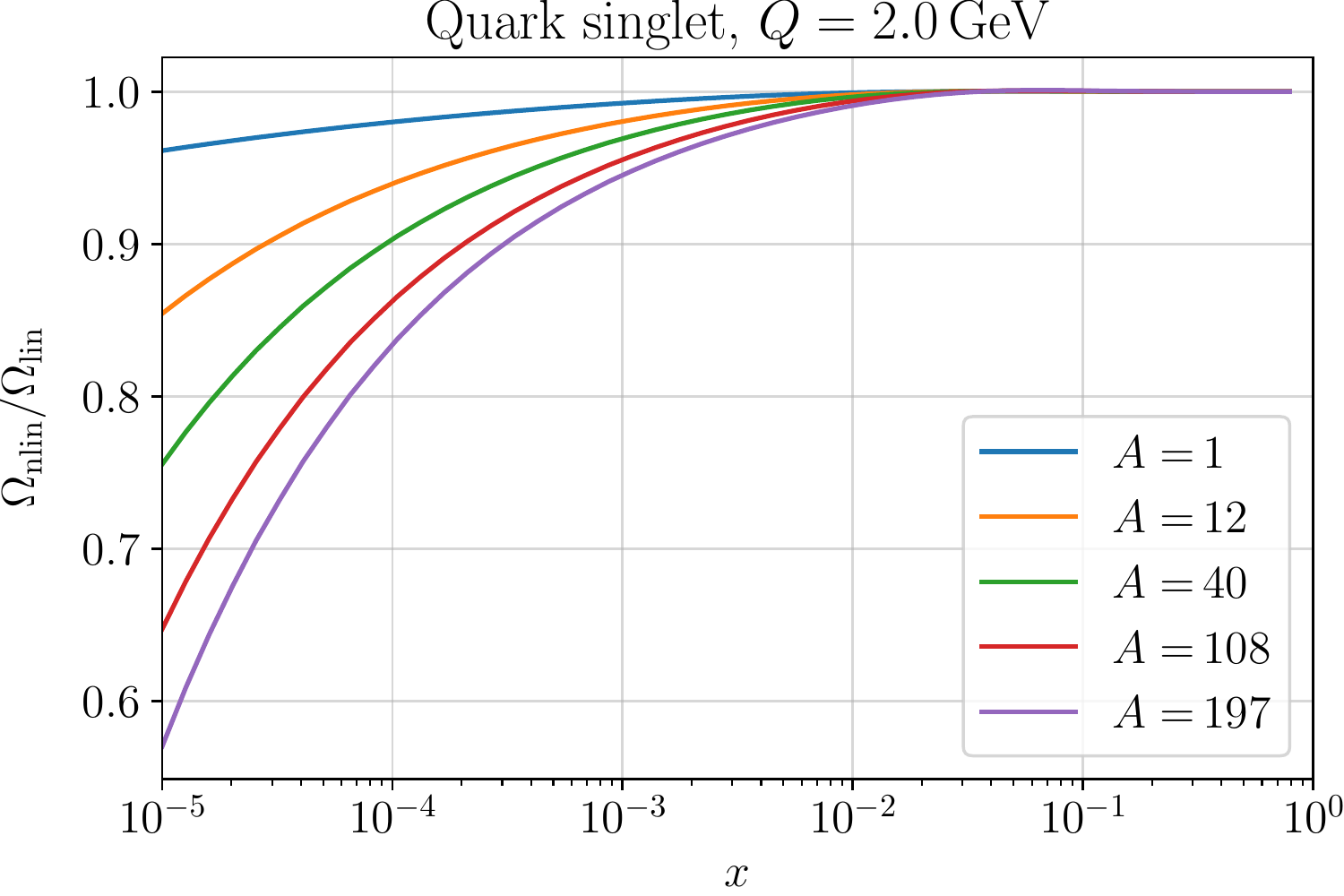,width=0.475\textwidth}
\epsfig{file=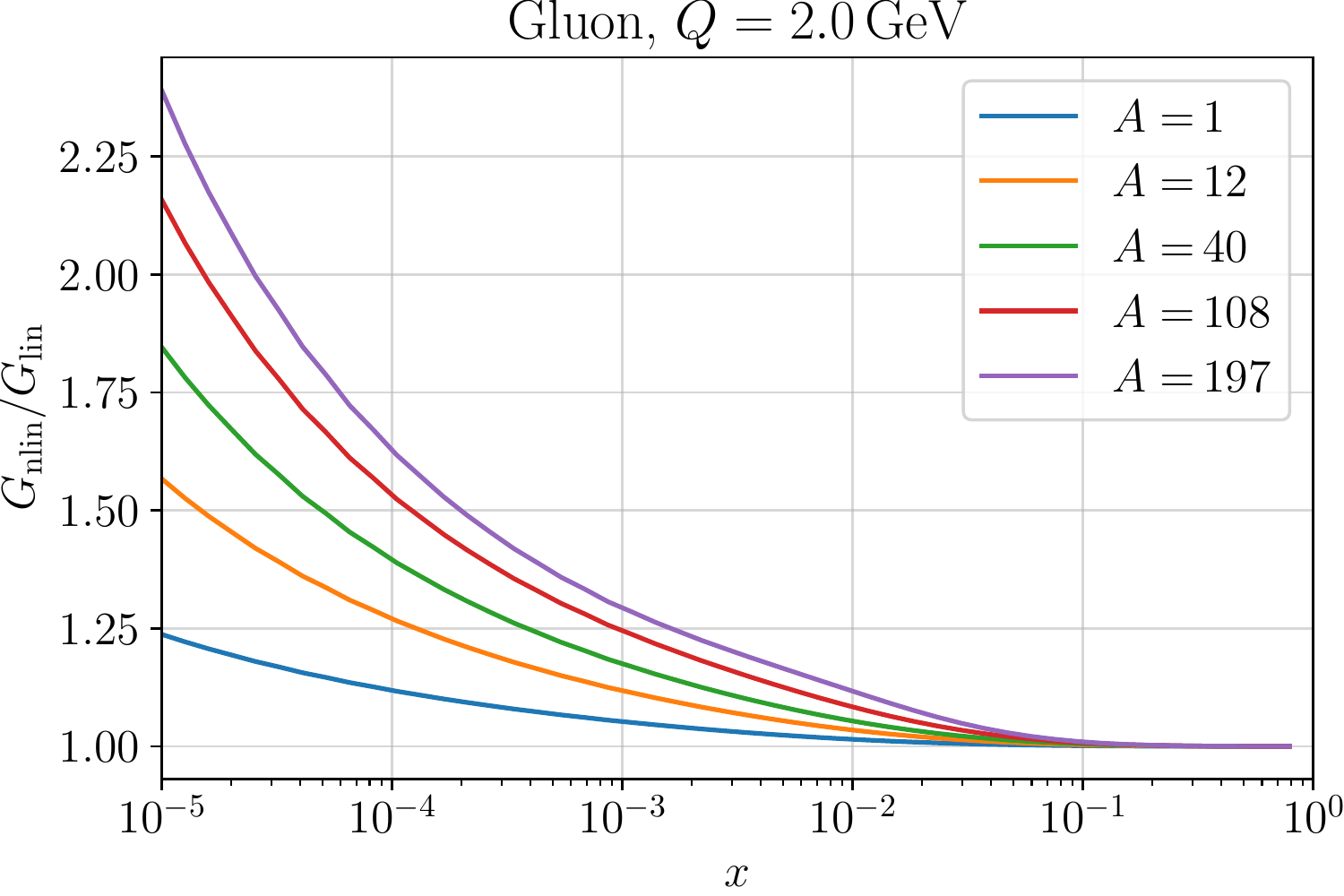,width=0.475\textwidth}
\caption{The ratios of the 
quark singlet $\Omega_{\rm nlin}(x,Q^2)/\Omega_{\rm lin}(x,Q^2)$ and gluon  $G_{\rm nlin}(x,Q^2)/G_{\rm lin}(x,Q^2)$ distributions after the GLR-MQ and DGLAP evolution, respectively, as a function of $x$ for a wide range of nuclei including  C-12, Ca-40, Ag-108, Au-197 and the free proton.
The upper panels correspond to the upward evolution from $Q_0=2$ GeV to $Q=10$ GeV; the lower panels are the results of the downward evolution from $Q_0=10$ GeV to $Q=2$ GeV. 
As input, the nCTEQ15 nPDFs have been used.
}
\label{fig:A_plots}
\end{figure}

Figure~\ref{fig:A_plots} quantifies the size of the nonlinear corrections as a function of the mass number $A$.
It presents the ratios of the
quark singlet and gluon distributions after the GLR-MQ and DGLAP evolution denoted by
$\Omega_{\rm nlin}(x,Q^2)/\Omega_{\rm lin}(x,Q^2)$ and $G_{\rm nlin}(x,Q^2)/G_{\rm lin}(x,Q^2)$, respectively,
as a function of $x$ for a wide range of nuclei including  C-12, Ca-40, Ag-108, Au-197 and the free proton.
As input, we used the nCTEQ15 nPDFs.
The two upper panels correspond to the result of the upward evolution from $Q_0=2$ GeV to $Q=10$ GeV; 
the two lower panels are the results of the downward evolution from $Q_0=10$ GeV to $Q=2$ GeV.
One can see from the figure that the nonlinear effects clearly become more important with increasing $A$. 
For example, at $x = 10^{-5}$, the upward evolution result for the proton gluon distribution is modified by about 4.5\%  
 and for the Au-197 distribution by about 15\%.
 The difference between the GLR-MQ and DGLAP evolved PDFs grows steadily with a decrease of $x$; it is largest at the smallest values of $x$ and disappears for $x \geq 0.1$.
 This behavior matches the approximate analytical solutions of the GLR-QM equations obtained by other groups~\cite{Boroun:2013mgv,Lalung:2018mpw,Devee:2014ida}. 

As in the case of Fig.~\ref{fig:nonlin_up_down}, we find that that the nonlinear terms have a much bigger relative impact on the downward evolution for all considered nuclei and the proton.
As explained previously, the nonlinear corrections suppress the quark singlet distribution and increase the gluon one.
For very small $x$ and heavy nuclei, the effect is ${\cal O}(30-40\%)$ for the quarks and ${\cal O}(100-140\%)$
for the gluons.

Using the obtained nuclear PDFs, one can readily calculate the NLO nuclear structure function $F_2^A(x,Q^2)$, 
\begin{eqnarray}
F_2^A(x,Q^2) &=& N(x,Q^2)+\frac{\alpha_s(Q^2)}{2 \pi} \int_x^{1} \frac{dz}{z^2} xC^{(1)}_{2q}\left(\frac{x}{z}\right) N(z,Q^2) 
\nonumber\\
&+& \langle e^2 \rangle \Omega(x,Q^2)+\langle e^2 \rangle \frac{\alpha_s(Q^2)}{2 \pi} \int_x^{1} \frac{dz}{z^2} xC^{(1)}_{2q}\left(\frac{x}{z}\right) \Omega(z,Q^2)+\langle e^2 \rangle \frac{\alpha_s(Q^2)}{2 \pi} \int_x^{1} \frac{dz}{z^2} xC^{(1)}_{2g}\left(\frac{x}{z}\right) G(z,Q^2),\   
\label{eq:F2A}
\end{eqnarray}
and the longitudinal structure function $F_L^A(x,Q^2)$,
\begin{eqnarray}
F^{A}_L(x,Q^2) &=& \frac{\alpha_s(Q^2)}{2 \pi} \int_x^{1} \frac{dz}{z^2} xC^{(1)}_{Lq}\left(\frac{x}{z}\right) N(z,Q^2) 
+\langle e^2 \rangle \frac{\alpha_s(Q^2)}{2 \pi} \int_x^{1} \frac{dz}{z^2} xC^{(1)}_{Lq}\left(\frac{x}{z}\right) \Omega(z,Q^2)
\nonumber\\
&+&\langle e^2 \rangle \frac{\alpha_s(Q^2)}{2 \pi} \int_x^{1} \frac{dz}{z^2} xC^{(1)}_{Lg}\left(\frac{x}{z}\right) G(z,Q^2)  \,.
\label{eq:F2L}
\end{eqnarray}
In Eqs.~(\ref{eq:F2A}) and (\ref{eq:F2L}), $N(x,Q^2)=x\sum_{i=1}^{N_F} e_i^2 q_i^{+}(x,Q^2)$ and 
$q_i^{+}=q_i(x,Q^2)+\bar{q}_i(x,Q^2)-1/N_F \Sigma(x,Q^2)$ are
non-singlet quark distributions with $N_F$ being the number of active flavors;  $\langle e^2 \rangle=(1/N_F) \sum_{i=1}^{N_F}
e_i^2$; $C^{(1)}_{2q}$, $C^{(1)}_{Lq}$, $C^{(1)}_{2g}$, and $C^{(1)}_{Lg}$ are the standard quark and gluon coefficient functions,
respectively.
The convolution integrals in Eqs.~(\ref{eq:F2A}) and (\ref{eq:F2L}) have exactly the same structure as those in the DGLAP 
evolution equations and, hence, the numerical method explained in Sec.~\ref{sec:method} can be used to evaluate them. 
Since the nonsinglet distribution $N(x,Q^2)$ is independent of $G(x,Q^2)$, we directly use the nCTEQ15 parametrization for it.

\begin{figure}[t]
\centering
\epsfig{file=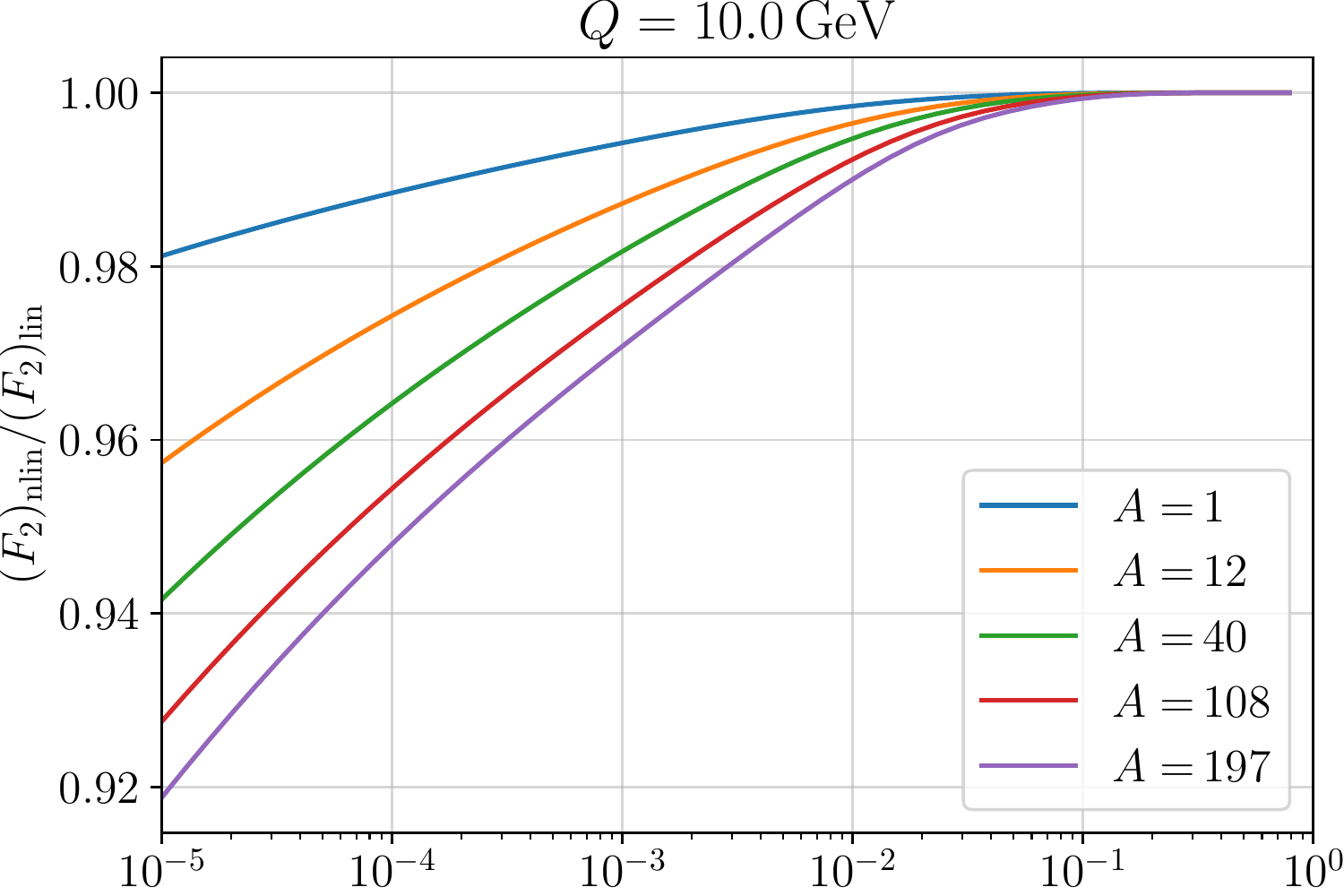,width=0.475\textwidth}
\epsfig{file=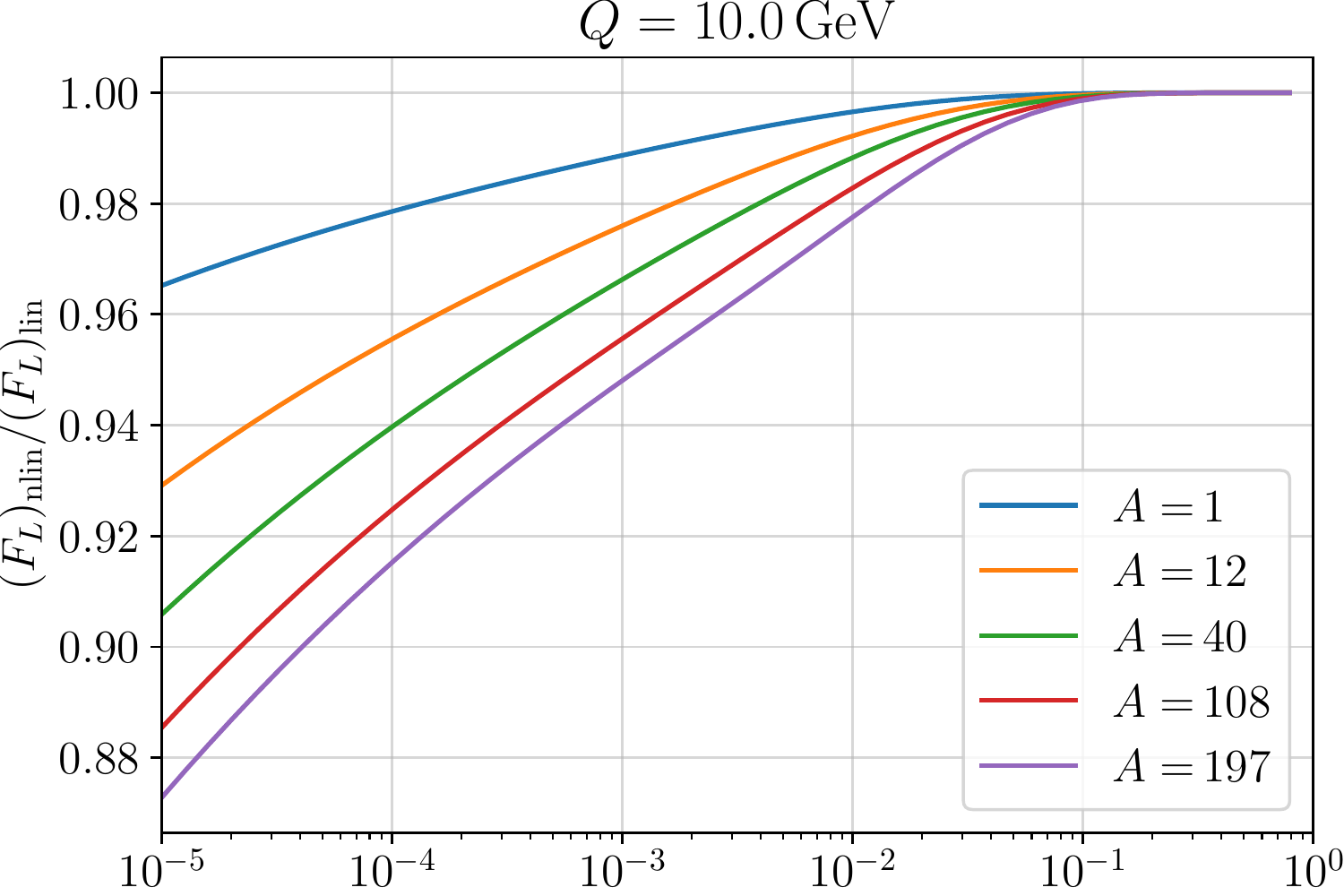,width=0.475\textwidth}
\epsfig{file=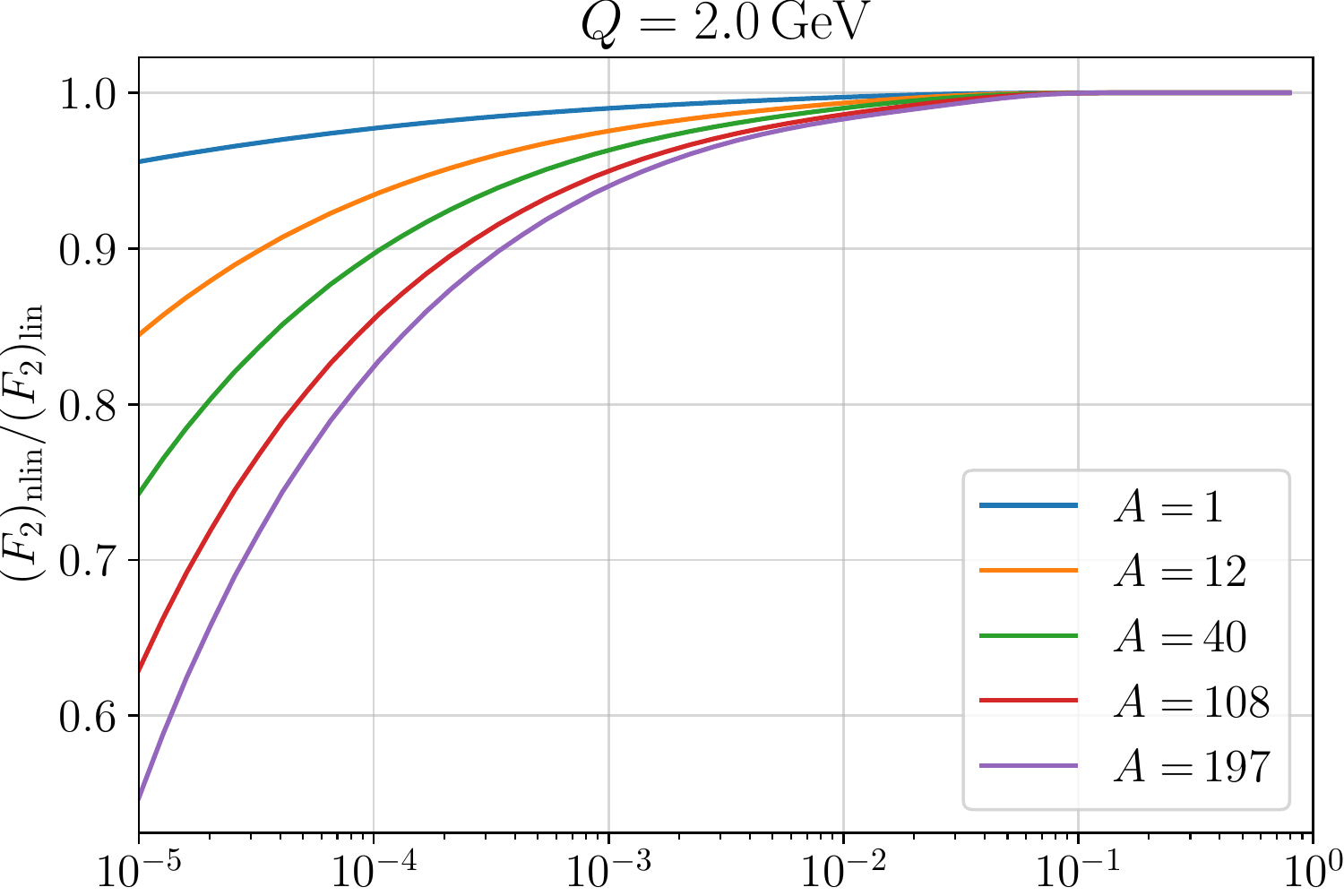,width=0.475\textwidth}
\epsfig{file=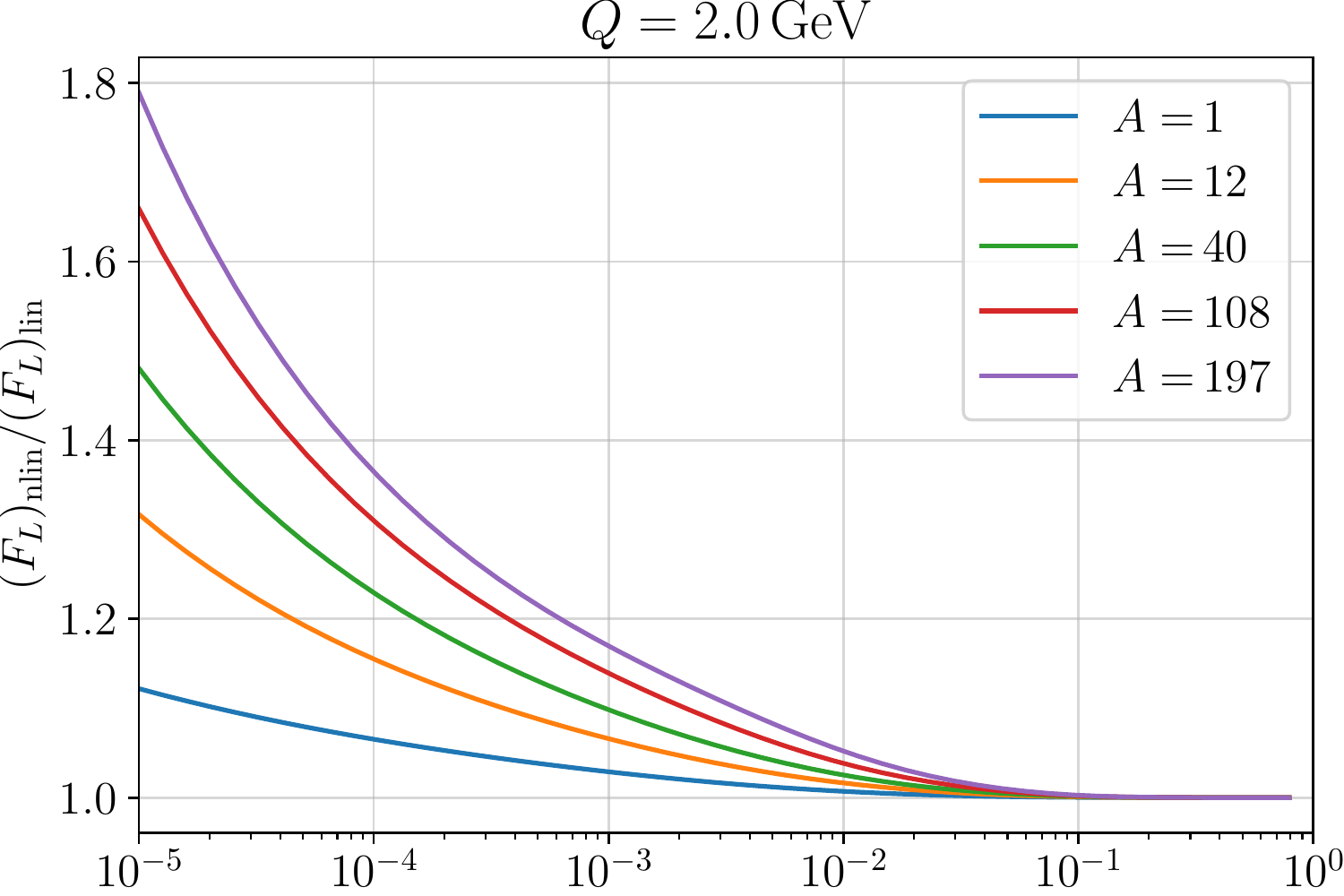,width=0.475\textwidth}
\caption{The ratios of the $F_2^A(x,Q^2)$ (left panels) and $F_L^A(x,Q^2)$ (right panels) structure functions
evaluated using the nuclear PDFs evolved according to the GLR-MQ and DGLAP evolution equations, respectively,
as a function of $x$. As in Fig.~\ref{fig:A_plots}, the upper panels correspond to the upward evolution from $Q_0=2$ GeV to $Q=10$ GeV, and the lower panels are for the downward evolution from $Q_0=10$ GeV to $Q=2$ GeV.
As input, the nCTEQ15 nPDFs have been used.
}
\label{fig:SF_A}
\end{figure}

Figure~\ref{fig:SF_A} shows the ratios of the $F_2^A(x,Q^2)$ (left panels) and $F_L^A(x,Q^2)$ (right panels) structure functions
evaluated using the nuclear PDFs, which were evolved according to the GLR-MQ and DGLAP evolution equations, respectively, employing the nCTEQ15 input. The ratios are denoted by $(F_2)_{\rm nlin}/(F_2)_{\rm lin}$ and $(F_L)_{\rm nlin}/(F_L)_{\rm lin}$ and are plotted as a function of $x$ for C-12, Ca-40, Ag-108, Au-197, and the free proton.
The trends of the $A$ and $x$ dependence mirror those of nPDFs shown in Fig.~\ref{fig:A_plots}, where 
$F_2^A(x,Q^2)$ is dominated by $\Omega(x,Q^2)$ and $F_{L}^A(x,Q^2)$ by $G(x,Q^2)$.
The nonlinear effects are again most important for heavy nuclei, and their impact is larger for 
$F_{L}^A(x,Q^2)$ than for $F_2^A(x,Q^2)$.
Thus, it should be easier to observe them experimentally by measuring $F_L^A(x,Q^2)$. For instance, when evolving upward, the structure function $F_L(x,Q^2)$ for the proton is modified by about 3.5\% at $x= 10^{-5}$, see the upper right panel. 
A similar-size effect can already be observed at $x =4 \times 10^{-3}$ for Au-197.

Note that the momentum sum rule for nPDFs is slightly violated in the GLM-MQ approach since the gluon-gluon recombination leads to a suppression of the singlet quark and gluon nPDFs after the upward evolution 
and to a suppression of the singlet quark and an enhancement of the gluon nPDFs after the downward evolution.
In particular, we find in the case of Au-197 that the total momentum sum rule is violated by approximately 5\% after the upward evolution and by 
less than 3\% after the downward evolution.

A generalization of this approach, 
which corrects this shortcoming and is valid in the whole $x$ region, was suggested~\cite{Zhu:1999ht}.
In our analysis, we focus only on the small $x$ region and, hence, do not address the issue of the momentum sum rule, which 
affects the picture of nuclear modifications of nPDFs, including the valence quarks, in a broad range of $x$.

\section{Conclusions}
\label{sec:conclusions}

In this paper, we numerically studied the GLR-MQ evolution equations for nPDFs to NLO accuracy and quantified the impact of gluon 
recombination at small $x$. Using the nCTEQ15 
and EPPS21
nPDFs as input, we confirmed the importance of the nonlinear corrections for small 
$x \lesssim 10^{-3}$, whose magnitude increases with a decrease of $x$ and an increase of the atomic number $A$.
For instance, at $x=10^{-5}$ and for heavy nuclei, after the upward evolution from $Q_0=2$ GeV to $Q=10$ GeV, the quark singlet 
$\Omega(x,Q^2)$ and the gluon $G(x,Q^2)$ distributions become reduced compared to the results of the nCTEQ15 parametrization
by $9-15$\%, respectively.
The relative effect is much stronger for the downward evolution from $Q_0=10$ GeV to $Q=2$ GeV, 
where we find that $\Omega(x,Q^2)$ is suppressed by 40\%, while $G(x, Q^2)$ is enhanced by 140\%. 
This is a consequence of the fact that the gluon-gluon recombination plays a much bigger role than the gluon-quark
splitting. 

The observed trend of the  behavior of nPDFs affects the $F_2^A(x,Q^2)$ and $F_L^A(x,Q^2)$ nuclear structure functions. In particular, we find that after the downward evolution from high to low $Q$ and for heavy nuclei and very small $x$, 
the $F_2^A(x,Q^2)$ structure function, which is dominated by $\Omega(x,Q^2)$, is reduced by 45\%, while 
the $F_L^A(x,Q^2)$ longitudinal structure function, which is predominantly sensitive to $G(x,Q^2)$, is enhanced by 80\%.
Our analysis indicates that the nonlinear effects are most pronounced in $F_L^A(x,Q^2)$ and are already quite sizable at
$x \sim 10^{-3}$ for heavy nuclei.
Since the results employing the EPPS21 nPDFs are quantitatively similar, our predictions very weakly depend on the choice of input nPDFs.

\acknowledgments

The work of V.G.\ is supported by the Academy of Finland (project No. 330448), the Centre of Excellence in Quark Matter (project No. 346326), 
the European Research Council project ERC-2018-ADG-835105 YoctoLHC, and also partially by a grant of Deutscher Akademischer Austauschdienst (DAAD). 
V.G.\ and J.R.\ would also like to thank Institut f\"ur Theoretische Physik, Westf\"alische Wilhelms-Universit\"at M\"unster for hospitality. M.K.\ thanks the School of Physics at the University of New South Wales in Sydney, Australia for its hospitality and financial support through the Gordon Godfrey visitors program. The work of M.K. was also funded by the DFG through grants KL 1266/9-1 and 10-1, the Research Training Group 2149 "Strong and Weak Interactions - from Hadrons to Dark Matter", and the SFB 1225 ``Isoquant'', project-id 273811115.

\end{document}